\documentclass[aps,preprint]{revtex4}
\usepackage{amssymb}
\usepackage{amsfonts}
\usepackage{amsmath}
\usepackage{graphicx}

\begin{document}

\title{Stability boundaries and collisions of two-dimensional solitons in $%
\mathcal{PT}$-symmetric couplers with the cubic-quintic nonlinearity}
\author{Gennadiy Burlak$^{1}$ and Boris A. Malomed$^{2}$}
\affiliation{$^{1}$Centro de Investigaci\'{o}n en Ingenier\'{\i}a y
Ciencias Aplicadas, Universidad Aut\'{o}noma del Estado de Morelos,
Cuernavaca, Mor., M\'{e}xico\\
$^{2}$Department of Physical Electronics, School of Electric
Engineering, Faculty of Engineering, Tel Aviv University, Tel Aviv
69978, Israel}

\begin{abstract}
We introduce one- and two-dimensional (1D and 2D) models of parity-time ($%
\mathcal{PT}$) -symmetric couplers with the mutually balanced linear gain
and loss applied to the two cores, and cubic-quintic (CQ) nonlinearity
acting in each one. The 2D and 1D models may be realized in dual-core
optical waveguides, in the spatiotemporal and spatial domains, respectively.
Stationary solutions for $\mathcal{PT}$-symmetric solitons in these systems
reduce to their counterparts in the usual coupler. The most essential
problem is the stability of the solitons, which become unstable against
symmetry breaking with the increase of the energy (norm), and retrieve the
stability at still larger energies. The boundary value of the
intercore-coupling constant, above which the solitons are completely stable,
is found by means of an analytical approximation, based on the CW
(zero-dimensional) counterpart of the system. The approximation demonstrates
good agreement with numerical findings for the 1D and 2D solitons. Numerical
results for the stability limits of the 2D solitons are obtained by means of
the computation of eigenvalues for small perturbations, and verified in
direct simulations. Although large parts of the solitons families are
unstable, the instability is quite weak. Collisions between 2D solitons in
the $\mathcal{PT}$-symmetric coupler are studied by means of simulations.
Outcomes of the collisions are inelastic but not destructive, as they do not
break the $\mathcal{PT}$ symmetry.
\end{abstract}

\maketitle

\section{Introduction and the setting}

Wave-propagation models of physical media are naturally separated into two
generic classes, conservative and dissipative. Recently, it was recognized
that a more particular species of $\mathcal{PT}$ (parity-time)-symmetric
systems may be identified at the boundary between these generic types \cite%
{Bender_review,special-issues,review}. They are represented by dissipative
quantum-mechanical models, and by classical waveguides subject to the
condition of spatial antisymmetry between separated gain and loss. While in
the quantum theory the $\mathcal{PT}$-symmetric models are subjects of
theoretical studies, the similarity of the quantum-mechanical Schr\"{o}%
dinger equation to the paraxial propagation equation in optics makes it
possible to implement this concept in real physical settings, as proposed
theoretically~\cite{Muga} and demonstrated experimentally \cite{experiment}
in a number of works. These possibilities have drawn a great deal of
interest to the wave propagation in $\mathcal{PT}$-symmetric systems \cite%
{special-issues}, especially in the presence of spatially periodic complex
potentials,~with even real and imaginary odd parts, as required by the $%
\mathcal{PT}$ symmetry \cite{PT_periodic,review}.

The ubiquitous occurrence of the Kerr nonlinearity in photonic media is an
incentive for studies of nonlinear realizations of the $\mathcal{PT}$
symmetry in optics, including $\mathcal{PT}$-symmetric solitons \cite%
{Musslimani2008} and their stability~\cite{Yang}. Dark solitons in $\mathcal{%
PT}$-symmetric systems were studied too, in the case of the self-defocusing
sign of the Kerr nonlinearity \cite{dark}, as well as bright solitons
supported by the quadratic nonlinearity \cite{chi2} and discrete solitons in
chains of coupled $\mathcal{PT}$-symmetric elements \cite%
{discrete,circular,KPZ}.

In contrast with the usual nonlinear systems including loss the gain terms,
where dissipative solitons exist as isolated \textit{attractors} \cite%
{PhysicaD,Kutz}, $\mathcal{PT}$-symmetric solitons emerge in continuous
families, similar to their counterparts in conservative media. However,
existence and stability domains for $\mathcal{PT}$-symmetric solitons shrink
with the increase of the gain-loss coefficient ($\gamma $) in the $\mathcal{%
PT}$-symmetric system, and they completely vanish at critical points, $%
\gamma =\gamma _{\max }$ and $\gamma =\gamma _{\mathrm{C}}$, as concerns the
existence and stability, respectively. As shown below, $\gamma _{\mathrm{C}}$
may be considerably smaller than $\gamma _{\max }$.

Following the addition of nonlinear terms to the conservative part of the $%
\mathcal{PT}$-symmetric system, its gain-and-loss part may also be made
nonlinear, by introducing balanced terms accounting for the cubic gain and
loss \cite{AKKZ}. Conditions for the existence and stability of optical
solitons under effects of combined linear and nonlinear $\mathcal{PT}$~terms
were addressed too \cite{combined}.

A model which is especially convenient for the studies of $\mathcal{PT}$%
-symmetric solitons is based on the coupler (dual-core system) with the
symmetric intrinsic Kerr nonlinearity, and the gain and loss applied
antisymmetrically to the two cores. This model was independently introduced
in Refs. \cite{Driben} and \cite{Canberra}, and extended, in various
directions, in works \cite{Driben2} and \cite{Konotop}. In the general form,
the model, which describes the spatiotemporal propagation of light in the
dual-core planar optical waveguide, is based on a system of two-dimensional
(2D) nonlinear Schr\"{o}dinger (NLS) equations for amplitudes of the
electromagnetic field in two cores, $\Psi $ and $\Phi $, coupled by the
linear terms, which account for the tunneling of light between the cores:%
\begin{equation}
i\frac{\partial \Psi }{\partial z}+\nabla ^{2}\Psi +N\left( \left\vert \Psi
\right\vert ^{2}\right) \Psi +\lambda \Phi =i\gamma \Psi ,  \label{Psi}
\end{equation}%
\begin{equation}
i\frac{\partial \Phi }{\partial z}+\nabla ^{2}\Phi +N\left( \left\vert \Phi
\right\vert ^{2}\right) \Phi +\lambda \Psi =-i\gamma \Phi ,  \label{Phi}
\end{equation}%
where $z$ is the propagation distance, $\nabla ^{2}\equiv \partial
^{2}/\partial x^{2}+\partial ^{2}/\partial y^{2}$ accounts for the
combination of the paraxial diffraction and anomalous group-velocity
dispersion, acting on the transverse coordinate $x$ and temporal variable $y$
in each core \cite{Dror:2011a} ($y$ is absent in the 1D version of the
model, which describes the operation of the dual-core waveguide in the
spatial domain), $N$ represents the intrinsic nonlinearity, $\lambda >0$ is
the coupling constant, and $\gamma $ is the above-mentioned balanced
gain-loss coefficient.

As shown in Ref. \cite{Driben}, stationary $\mathcal{PT}$-symmetric
solutions (including solitons) with propagation constant $k$ can be found in
a generic form,%
\begin{equation}
\Psi \left( z,x,y\right) =e^{ikz-i\delta /2}U\left( x,y\right) ,~\Phi \left(
z,x,y\right) =e^{ikz+i\delta /2}U\left( x,y\right) ,  \label{U}
\end{equation}%
where the constant phase shift between the components is%
\begin{equation}
\delta =\arcsin \left( \gamma /\lambda \right) ,  \label{delta}
\end{equation}%
and real function $U$ satisfies the usual stationary NLS equation, with a
shifted value of the propagation constant:%
\begin{equation}
-\left( k-\lambda _{\mathcal{PT}}\right) U+\nabla ^{2}U+N\left( U^{2}\right)
U=0,  \label{UU}
\end{equation}%
\begin{equation}
\lambda _{\mathcal{PT}}\equiv \sqrt{\lambda ^{2}-\gamma ^{2}}.  \label{PT}
\end{equation}%
Obviously, these solutions exist under condition%
\begin{equation}
\gamma \leq \gamma _{\max }\equiv \lambda ,  \label{cr}
\end{equation}%
which determines the above-mentioned largest value of the gain-loss
coefficient for the $\mathcal{PT}$-symmetric coupler. Localized solutions to
Eq. (\ref{UU}), i.e., solitons, are possible for $k>$ $\lambda _{\mathcal{PT}%
}$.

Because the actual transverse width of the waveguide, $X$, is finite,
solitons are meaningful solutions if their size in the $x$-direction is
essentially smaller than $X$. Length $Z$ of the waveguide in the
longitudinal direction is finite too, which implies that the soliton
solutions are relevant ones if their diffraction and dispersion lengths are
much smaller than $Z$. These conditions definitely hold in the analysis
presented below.

While the shift of the effective coupling constant, given by Eq. (\ref{PT}),
is an obvious result, a crucial issue is the stability of the $\mathcal{PT}$%
-symmetric solitons in the couplers against symmetry-breaking perturbations.
In the usual conservative models, symmetric solitons in couplers with the
Kerr \cite{Wabnitz,coupler-Kerr} or quadratic \cite{coupler-chi2} intrinsic
nonlinearity become unstable at a critical value of the total energy, alias
norm,%
\begin{equation}
E=\int \int \left[ \left\vert \Psi \left( x,y\right) \right\vert
^{2}+\left\vert \Phi \left( x,y\right) \right\vert ^{2}\right] dxdy\equiv
E_{\Psi }+E_{\Phi },  \label{E}
\end{equation}%
$E=E_{\mathrm{C}}^{(\mathrm{coupler})}(\lambda ),$ and at $E>E_{\mathrm{C}%
}^{(\mathrm{coupler})}(\lambda )$ unstable symmetric solitons are replaced
by stable asymmetric ones, which is a typical manifestation of the \textit{%
spontaneous symmetry breaking} \cite{book}.

As shown in Ref. \cite{Driben}, the stability boundary for symmetric
solitons in $\mathcal{PT}$-symmetric couplers can be obtained by replacing
the coupling constant, $\lambda $, by its effective value (\ref{PT}),
\begin{equation}
E_{\mathrm{C}}^{(\mathcal{PT})}\left( \lambda ,\gamma \right) =E_{\mathrm{C}%
}^{(\mathrm{coupler})}\left( \sqrt{\lambda ^{2}-\gamma ^{2}}\right) .
\label{Ecr}
\end{equation}%
In particular, for the 1D $\mathcal{PT}$-symmetric coupler with the cubic
nonlinearity, the stability boundary was found in Refs. \cite{Driben} and
\cite{Canberra} in an exact form, making use of the fact that $E_{\mathrm{C}%
}^{(\mathrm{coupler})}(\lambda )$ is available in an exact form in the model
of the usual coupler with the Kerr nonlinearity \cite{Wabnitz}, see more
details below. However, a drastic difference of the $\mathcal{PT}$-symmetric
coupler from its conservative counterpart is that, beyond the
symmetry-breaking boundary, unstable symmetric solitons are not replaced by
asymmetric modes, but rather blow up. Indeed, an obvious corollary of Eqs. (%
\ref{Psi}), (\ref{Phi}) and (\ref{E}) is the energy-balance equation,%
\begin{equation}
\frac{dE}{dz}=2\gamma \left( E_{\psi }-E_{\Phi }\right) ,  \label{dE/dz}
\end{equation}%
hence only symmetric solitons, with $E_{\psi }=E_{\Phi }$, may represent
stationary modes. A rigorous proof of the nonexistence of asymmetric
solitons in $\mathcal{PT}$-symmetric systems was recently presented in Ref.
\cite{JYang}.

The main objective of the present work is to find stability limits for
fundamental solitons in two-dimensional (2D) $\mathcal{PT}$-symmetric
couplers, which, as mentioned above, may be realized as planar dual-core
optical waveguides operating in the spatiotemporal domain \cite{Dror:2011a},
with the gain and loss applied to the two cores. To avoid the collapse in
the 2D setting, driven by the cubic self-focusing nonlinearity \cite%
{collapse}, it is necessary to include self-defocusing quintic terms acting
in each core \cite{review-Wise}. The combined cubic-quintic (CQ)
nonlinearity of this type occurs in various optical media \cite{CQ}. The
competition of the cubic and quintic nonlinearities makes the spontaneous
symmetry breaking of solitons in the respective usual (non-$\mathcal{PT}$)
coupler drastically different from the situation in the case of the cubic
self-focusing: the symmetric solitons are unstable, and asymmetric solitons
exist, in \emph{finite intervals} of energies, as shown by means of
numerical methods in the 1D \cite{Albuch,Zeev} and 2D \cite{Dror:2011a}
versions of the system (recently, a similar result was demonstrated for the
1D coupler with competing quadratic and cubic nonlinearities \cite{Lazar}).
The width of the intervals depends in the inter-core coupling constant, $%
\lambda $ [see Eqs. (\ref{GPE-U}) and (\ref{GPE-V}) below], shrinking to nil
and vanishing at some value $\lambda _{\max }$. An analytical estimate for $%
\lambda _{\max }$, including its modification for the $\mathcal{PT}$%
-symmetric coupler, is obtained below in Section II, see Eqs. (\ref{1/8})
and (\ref{largest}) (previously, $\lambda _{\max }$ was found in a numerical
form only, even in the absence of the gain and loss terms). Numerical
results for the stability of the 2D solitons, which are the most essential
findings reported in the present work, are presented in Section III. In the
same section, we report results of systematic simulations of collisions
between 2D solitons, which is an obviously relevant problem too. It is found
that collisions spontaneously break the symmetry between colliding identical
solitons, but do not break the $\mathcal{PT}$ symmetry and do not destroy
the solitons. The paper is concluded by Section IV.

\section{Analytical considerations}

The system of Eqs. (\ref{Psi}) and (\ref{Phi}) with the normalized CQ
nonlinearity, $N(\left\vert \Psi \right\vert ^{2})=\left\vert \Psi
\right\vert ^{2}-\left\vert \Psi \right\vert ^{4}$, is written in the
following scaled form:
\begin{equation}
i\frac{\partial \Psi }{\partial z}+\left( \frac{\partial ^{2}}{\partial x^{2}%
}+\frac{\partial ^{2}}{\partial y^{2}}\right) \Psi +\left\vert \Psi
\right\vert ^{2}\Psi -\left\vert \Psi \right\vert ^{4}\Psi +\lambda \Phi
=i\gamma \Psi ,  \label{GPE-U}
\end{equation}%
\begin{equation}
i\frac{\partial \Phi }{\partial z}+\left( \frac{\partial ^{2}}{\partial x^{2}%
}+\frac{\partial ^{2}}{\partial y^{2}}\right) \Phi +\left\vert \Phi
\right\vert ^{2}\Phi -\left\vert \Phi \right\vert ^{4}\Phi +\lambda \Psi
=-i\gamma \Phi .  \label{GPE-V}
\end{equation}%
The first objective is to construct families of fundamental solitons, i.e.,
localized ground-state modes, in the form of Eq. (\ref{U}), with
axisymmetric function $U\left( r\equiv \sqrt{x^{2}+y^{2}}\right) $
satisfying the ordinary differential equation,
\begin{equation}
-\left( k-\lambda _{\mathcal{PT}}\right) U+\frac{d^{2}U}{dr^{2}}+\frac{D-1}{r%
}\frac{dU}{dr}+U^{3}-U^{5}=0,  \label{D}
\end{equation}%
where $D=2$ or $1$ is the transverse dimension (in the 1D case, $r$ is
replaced by $x$); recall that $\lambda _{\mathcal{PT}}$ is defined as per
Eq. (\ref{PT}). The fundamental-soliton solutions to Eq. (\ref{D}) satisfy
the corresponding boundary conditions, $dU/dr(r=0)=0$, $U(r)\sim \exp \left(
-\sqrt{k-\lambda _{\mathcal{PT}}}r\right) $ at $r\rightarrow \infty $.

On the other hand, in the usual coupler model, with $\gamma =0$, which
admits not only symmetric but also asymmetric soliton modes, the stationary
solutions are looked for as $\left\{ \Psi \left( r,z\right) ,\Phi \left(
r,z\right) \right\} =e^{ikz}\left\{ U(r),V(r)\right\} ,$ with real functions
$U$ and $V$ satisfying coupled equations
\begin{equation}
kU=\frac{d^{2}U}{dr^{2}}+\frac{D-1}{r}\frac{dU}{dr}+U^{3}-U^{5}+\lambda V,
\label{kU}
\end{equation}%
\begin{equation}
kV=\frac{d^{2}V}{dr^{2}}+\frac{D-1}{r}\frac{dV}{dr}+V^{3}-V^{5}+\lambda U.
\label{kV}
\end{equation}%
The breaking and restoration of the symmetry of solitons is signalled by the
existence of a \textit{zero mode} of infinitesimal \emph{antisymmetric}
perturbations, $\left\{ U(x),V(x)\right\} =\left\{ U_{\mathrm{sol}}(x)\pm
\delta U(x)\right\} $, which satisfies the linear Schr\"{o}dinger equation
obtained by the subtraction of the linearized version of Eq. (\ref{kV}) from
its counterpart corresponding to Eq. (\ref{kU}):%
\begin{equation}
-\left( k+\lambda \right) \delta U=-\left( \frac{d^{2}}{dr^{2}}+\frac{D-1}{r}%
\frac{d}{dr}\right) \delta U+W^{\mathrm{(eff)}}(x)\delta U,  \label{Schr}
\end{equation}%
where $-\left( k+\lambda \right) $ plays the role of the energy eigenvalue
in the linear Schr\"{o}dinger equation, with effective potential%
\begin{equation}
W^{\mathrm{(eff)}}(x)=-3U_{\mathrm{sol}}^{2}(x)+5U_{\mathrm{sol}}^{4}(x).
\label{Weff}
\end{equation}%
In the 1D model with the cubic nonlinearity, Eq. (\ref{Schr}) admits an
exact solution, which makes it possible to find the respective exact
symmetry-breaking point \cite{Wabnitz}. The exact solution was extended for
the 1D $\mathcal{PT}$-symmetric coupler with the cubic nonlinearity in Refs.
\cite{Driben} and \cite{Canberra}.

The well-known exact soliton solution to the 1D version of the CQ equation (%
\ref{D}) is \cite{Bulgaria}%
\begin{equation}
U_{\mathrm{sol}}^{2}(x)=\frac{q}{1+\sqrt{1-\left( 4/3\right) q}\cosh \left(
\sqrt{q}x\right) },~q\equiv 4\left( k-\lambda _{\mathcal{PT}}\right) ,
\label{sol}
\end{equation}%
which exists for $0<q<3/4$. The norm of this soliton, defined as per the 1D
version of Eq. (\ref{E}), is%
\begin{equation}
E_{\mathrm{sol}}(q)=\sqrt{3}\ln \left( \frac{\sqrt{3}+2\sqrt{q}}{\sqrt{3}-2%
\sqrt{q}}\right) .  \label{E_sol}
\end{equation}

The 2D counterpart of soliton (\ref{sol}) can be found in a numerical form
\cite{Sweden,surface-tension}. Unlike the energy of the 1D solitons, which
starts from $E=0$ at $q=0$, i.e., at $k=\lambda _{\mathcal{PT}}$ [see Eqs. (%
\ref{E_sol}) and (\ref{sol})], the energy of the family of the 2D solitons
takes values above the threshold value: $E_{\mathrm{sol}}\geq E_{\mathrm{%
Townes}}\approx $ $\allowbreak 11.\,\allowbreak 69$, the threshold being the
commonly known energy of the (unstable) Townes solitons in the 2D NLS
equation with the cubic nonlinearity \cite{collapse}. In the limit of $q=3/4$%
, both the 1D and 2D solitons degenerate into the constant (CW,
continuous-wave) solution, with%
\begin{equation}
U^{2}=U_{\max }^{2}\equiv 3/4  \label{4/3}
\end{equation}%
and $E=\infty $ [CW solutions, $U_{\mathrm{CW}}^{2}=(1/2)\left( 1\pm \sqrt{%
1-q}\right) $, exist also at $3/4<q\leq 1$].

As mentioned above, in the model with the CQ nonlinearity 1D and 2D
asymmetric solitons exist in a finite interval of energies, $E_{\min
}(\lambda )<E<E_{\max }(\lambda )$, which shrinks to nil [$E_{\max }(\lambda
)-E_{\min }(\lambda )\rightarrow 0$] at
\begin{equation}
\lambda =\left\{
\begin{array}{c}
\lambda _{\max }^{(\mathrm{1D})}\approx 0.11, \\
\lambda _{\max }^{(\mathrm{2D})}\approx 0.096,%
\end{array}%
\right.   \label{lambda}
\end{equation}%
as found by means of numerical calculations in Refs. \cite{Albuch} and \cite%
{Dror:2011a} for the 1D and 2D systems, respectively. An analytical estimate
for $\lambda _{\max }$ can be obtained replacing Eqs. (\ref{kU}) and (\ref%
{kV}) by algebraic equations for the CW, i.e., zero-dimensional, states,
neglecting the derivatives in these equations. The accordingly simplified
Eq. (\ref{Schr}) for the zero symmetry-breaking mode of infinitesimal
perturbations reduces to relation
\begin{equation}
-\left( k+\lambda \right) =-3U^{2}+5U^{5}.  \label{CW}
\end{equation}%
At the symmetry-breaking and restoration points, Eq. (\ref{CW}) must hold
simultaneously with the CW version of Eqs. (\ref{kU}) and (\ref{kV}) for the
symmetric CW states, $U=V$, i.e.,
\begin{equation}
\left( k-\lambda \right) =U^{2}-U^{4}.  \label{CWsymm}
\end{equation}%
It then immediately follows from Eqs. (\ref{CW}) and (\ref{CWsymm}) that the
symmetry breaking ($-$) and restoration ($+$) of the CW states take place at
the following values of the amplitude and propagation constant:%
\begin{equation}
U_{0}^{2}=\frac{1}{4}\left( 1\pm \sqrt{1-8\lambda }\right) ,  \label{pm}
\end{equation}%
\begin{equation}
k_{0}=\frac{1}{8}\left( 1+12\lambda \pm \sqrt{1-8\lambda }\right) .
\label{kcr}
\end{equation}

These results predict the largest value of the coupling constant up to which
the symmetry breaking occurs for the CW states in the coupler's model,%
\begin{equation}
\lambda _{\max }^{\mathrm{(CW)}}=1/8,  \label{1/8}
\end{equation}%
which is reasonably close to the numerical values (\ref{lambda}) previously
found for the 1D and 2D solitons. The fact that $\lambda _{\max }$ is
somewhat smaller for solitons than for the CW states can be understood too,
because, whilst the central portion of the soliton fields may get into the
interval of%
\begin{equation}
\frac{1}{4}\left( 1-\sqrt{1-8\lambda }\right) <U^{2}<\frac{1}{4}\left( 1+%
\sqrt{1-8\lambda }\right) ,  \label{unst}
\end{equation}%
where the symmetric states are unstable, see Eq. (\ref{pm}), their decaying
tails fall into the range where no symmetry breaking occurs, thus pulling
the symmetric solitons closer to the stability range. Note that, for very
broad 1D and 2D solitons, the squared amplitude $U^{2}=3/4$ of the
quasi-flat segment [see Eq. (\ref{4/3})] does not fall into the instability
interval (\ref{unst}), i.e., such limit-form solitons exist only in the
symmetric form and are stable.

Lastly, it follows from Eq. (\ref{PT}) that, in the $\mathcal{PT}$-symmetric
system with the gain-loss coefficient, $0<\gamma <\lambda $, the region of
the \textit{absolute stability} of the symmetric solitons, which is $\lambda
>\lambda _{\max }^{(\mathrm{1D,2D}\text{)}}$ in the usual coupler, and is
approximated in the analytical form by Eq. (\ref{1/8}), is shifted to larger
values of the coupling constant:%
\begin{equation}
\lambda ^{2}>\lambda _{\mathcal{PT}}^{2}\equiv \left( \lambda _{\max }^{%
\mathrm{(1D,2D}\text{)}}\right) ^{2}+\gamma ^{2}.  \label{largest}
\end{equation}

\section{Numerical results for 2D solitons}

\subsection{Stationary solitons}

General relations (\ref{U}) reduce the construction of stationary 2D $%
\mathcal{PT}$-symmetric solitons to a numerical solution of Eq. (\ref{D}).
For this reason, the family of the fundamental solitons depends on the
single combination of parameters,
\begin{equation}
k_{\mathrm{eff}}\equiv k-\sqrt{\lambda ^{2}-\gamma ^{2}},  \label{keff}
\end{equation}%
see Eq. (\ref{PT}). A set of radial shapes of the 2D solitons (which are, in
principle, known from Ref. \cite{Sweden}) is displayed in Figs. \ref%
{Pic_Fig1} for fixed values $k=0.12$ and $\lambda =0.08$ [the latter one
makes sense, as it is smaller than $\lambda _{\max }^{\mathrm{(2D)}}$
defined in Eq. (\ref{lambda})] and $\gamma $ varying from $0$ up to the
maximum value, $\gamma _{\max }=\lambda $, beyond which the $\mathcal{PT}$%
-symmetric solitons do not exist [see Eq. (\ref{cr})]. The numerical method
used to generate these stationary solutions is based on the Newton-Raphson
iterations, implemented in the Cartesian coordinates \cite{CppRecipes2002}.
The solutions were obtained with relative accuracy $10^{-8}$.

\begin{figure}[tbp]
\begin{center}
\includegraphics[scale=0.80]{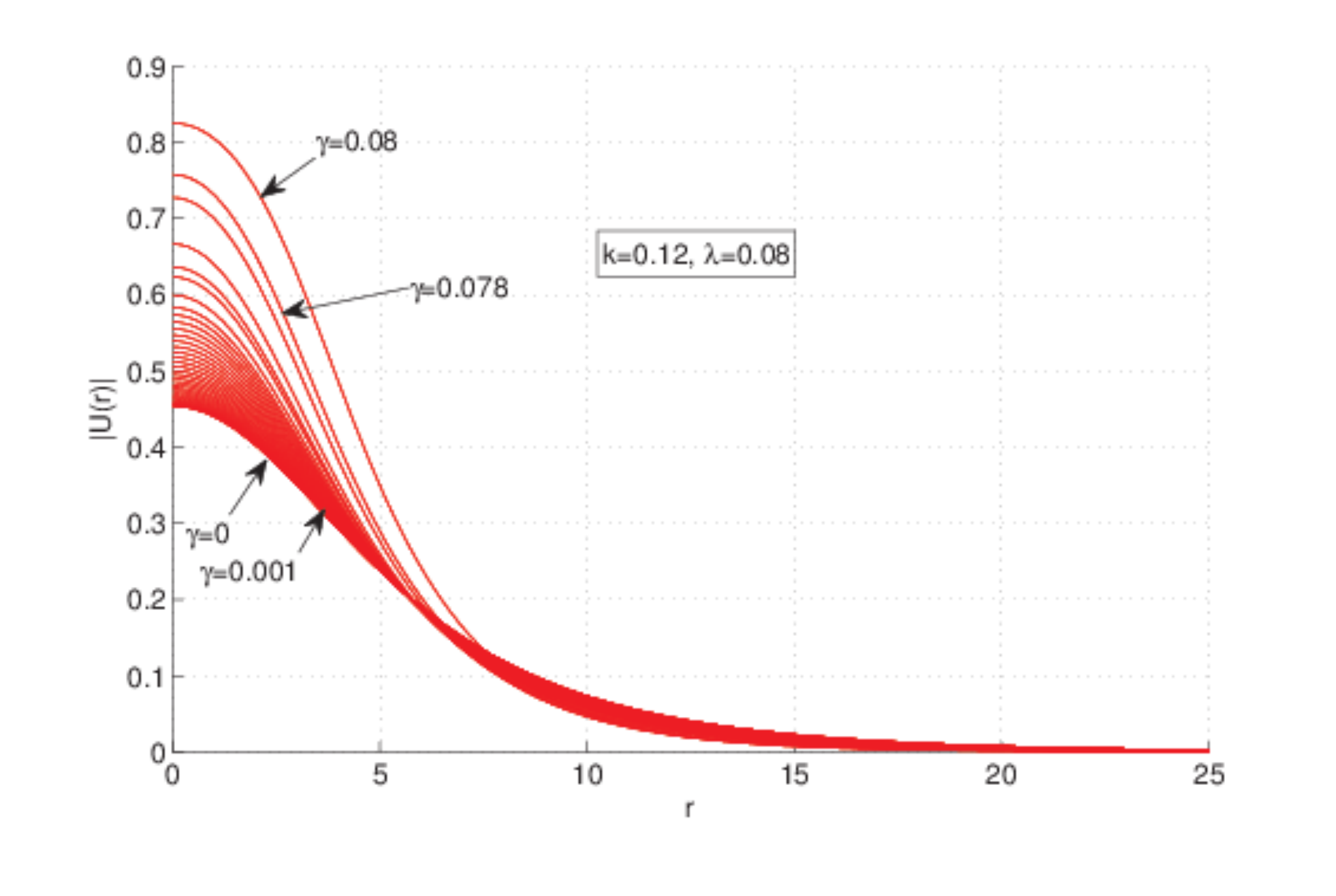}
\end{center}
\caption{(Color on line) A set of radial profiles of the 2D $\mathcal{PT}$%
-symmetric solitons, found in the numerical form at a fixed propagation
constant, $k=0.12$, and coupling constant $\protect\lambda =0.08$, for
various values of the gain-loss parameter, $\protect\gamma =0,0.001...,0.08$%
. }
\label{Pic_Fig1}
\end{figure}

\begin{figure}[tbp]
\begin{center}
\includegraphics[scale=0.80]{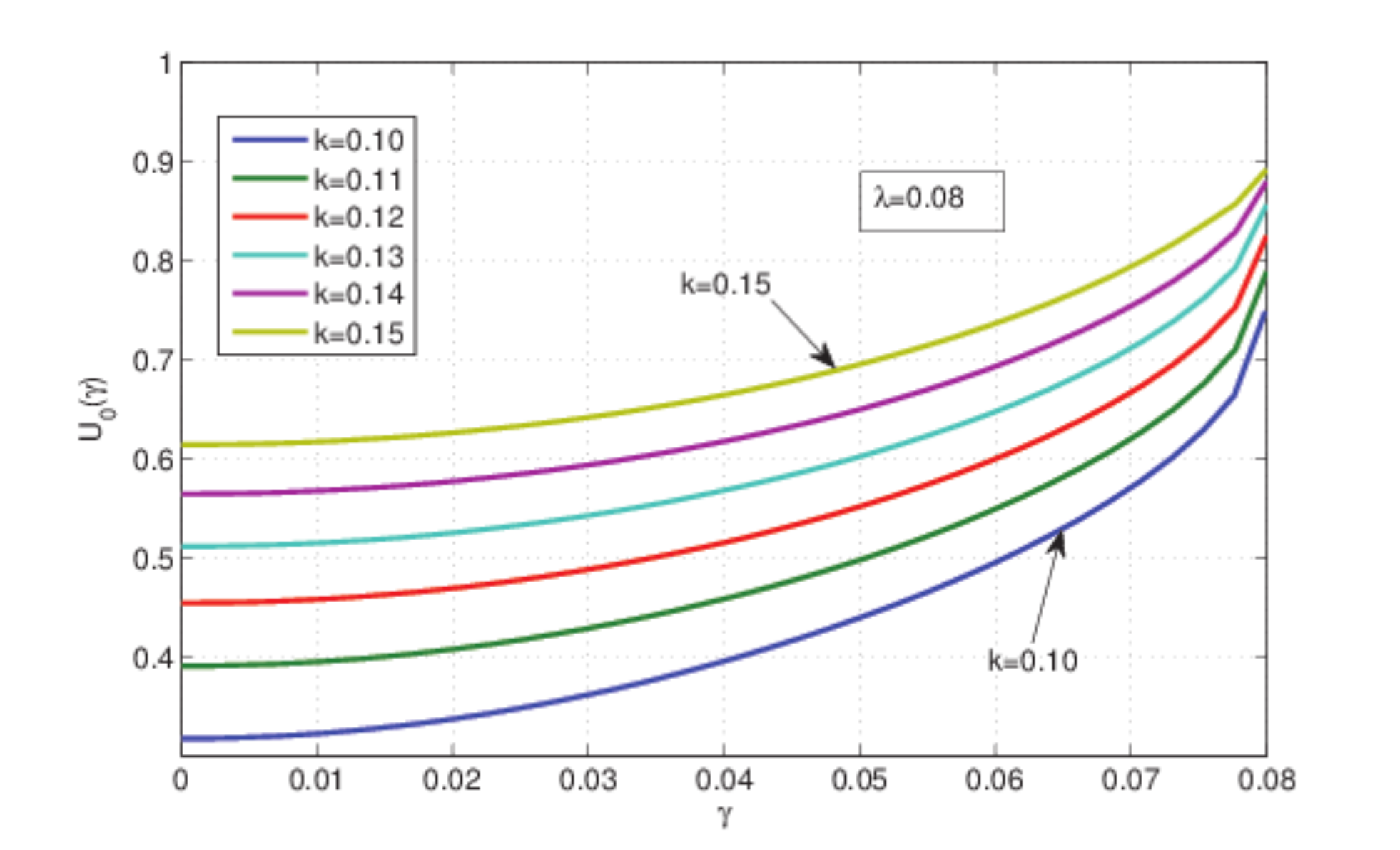}
\end{center}
\caption{(Color on line). The soliton's amplitude, $U(r=0)\equiv U_{0}(%
\protect\gamma )$, as a function of $\protect\gamma $ for a fixed coupling
constant, $\protect\lambda =0.08$,\ and different propagation constants $k$.}
\label{Pic_Fig2}
\end{figure}

Figure \ref{Pic_Fig1} shows that the soliton's amplitude, $U(r=0)$,
increases with the growth of $\gamma $, in accordance to the fact that
combination (\ref{keff}) increases with $\gamma $. The dependence of the
amplitude on $\gamma $ and $k$, for the same coupling constant as in Fig. %
\ref{Pic_Fig1}, $\lambda =0.08$, is presented in Fig. \ref{Pic_Fig2}
[strictly speaking, the amplitude also depends on the single combination $k_{%
\mathrm{eff}}$, see Eq. (\ref{keff}), but it makes sense to display the
dependence of the amplitude on $\gamma $ for various constant values of $k$).

\subsection{Stability limits for the solitons}

As mentioned above, Eq. (\ref{U}) essentially reduces the shape of the 2D
fundamental $\mathcal{PT}$-symmetric solitons to that which was found, in
another context, in Ref. \cite{Sweden}. A new issue in the context of the $%
\mathcal{PT}$ symmetry is the stability of the solitons, beneath boundary (%
\ref{largest}) [recall that the solitons are completely stable above it,
with $\lambda _{\max }^{\mathrm{(2D)}}$ found in the numerical form as
indicated in Ref. (\ref{lambda}), or approximated analytically as per Eq. (%
\ref{1/8})].

\begin{figure}[tbp]
\begin{center}
\includegraphics[scale=0.75]{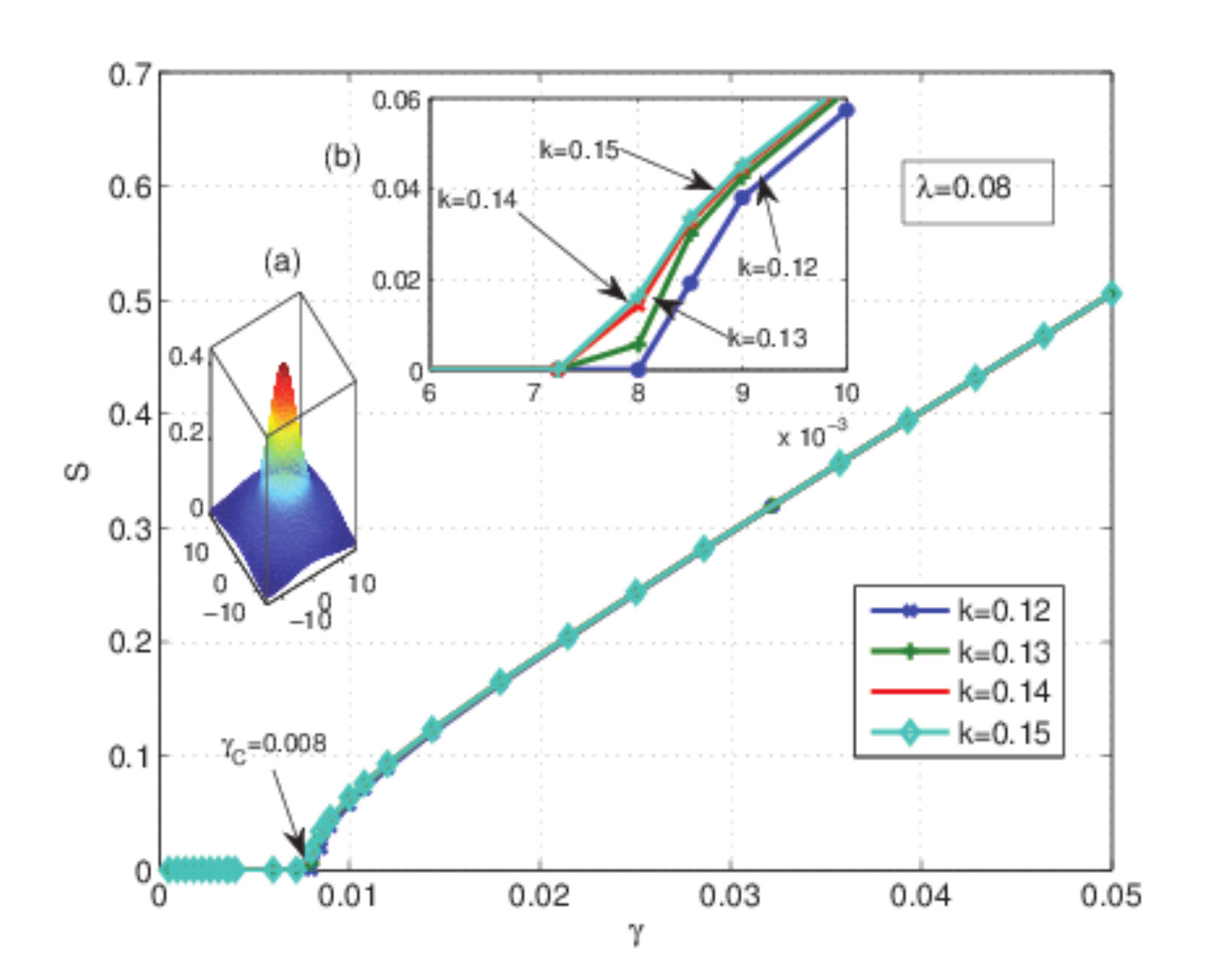}
\end{center}
\caption{(Color online) The largest instability growth rate for eigenmodes
of small perturbations around the $\mathcal{PT}$-symmetric 2D solitons, $S(%
\protect\gamma )=\max \left\{ \mathrm{Re}(\protect\sigma )\right\} \geq 0$ [$%
\protect\sigma $ is a complex eigenvalue of the linear-stability spectrum],
for the fixed coupling constant, $\protect\lambda =0.08$, and several values
of the propagation constant, $k=0.12$, $0.13$, $0.14$, $0.15$, versus the
gain-loss coefficient, $\protect\gamma $. Inset (a) shows the profile of a
typical stable soliton, found at $k=0.12$ and $\protect\gamma =0.002$. Inset
(b) is a blowup of a vicinity of the destabilization transition. For all
these values of $k$, the destabilization occurs around the critical value, $%
\protect\gamma _{\mathrm{C}}\approx 0.008$. The solitons are stable in the
region of $S(\protect\gamma )=0$.}
\label{Pic_Fig3}
\end{figure}

\begin{figure}[tbp]
\begin{center}
\includegraphics[scale=0.75]{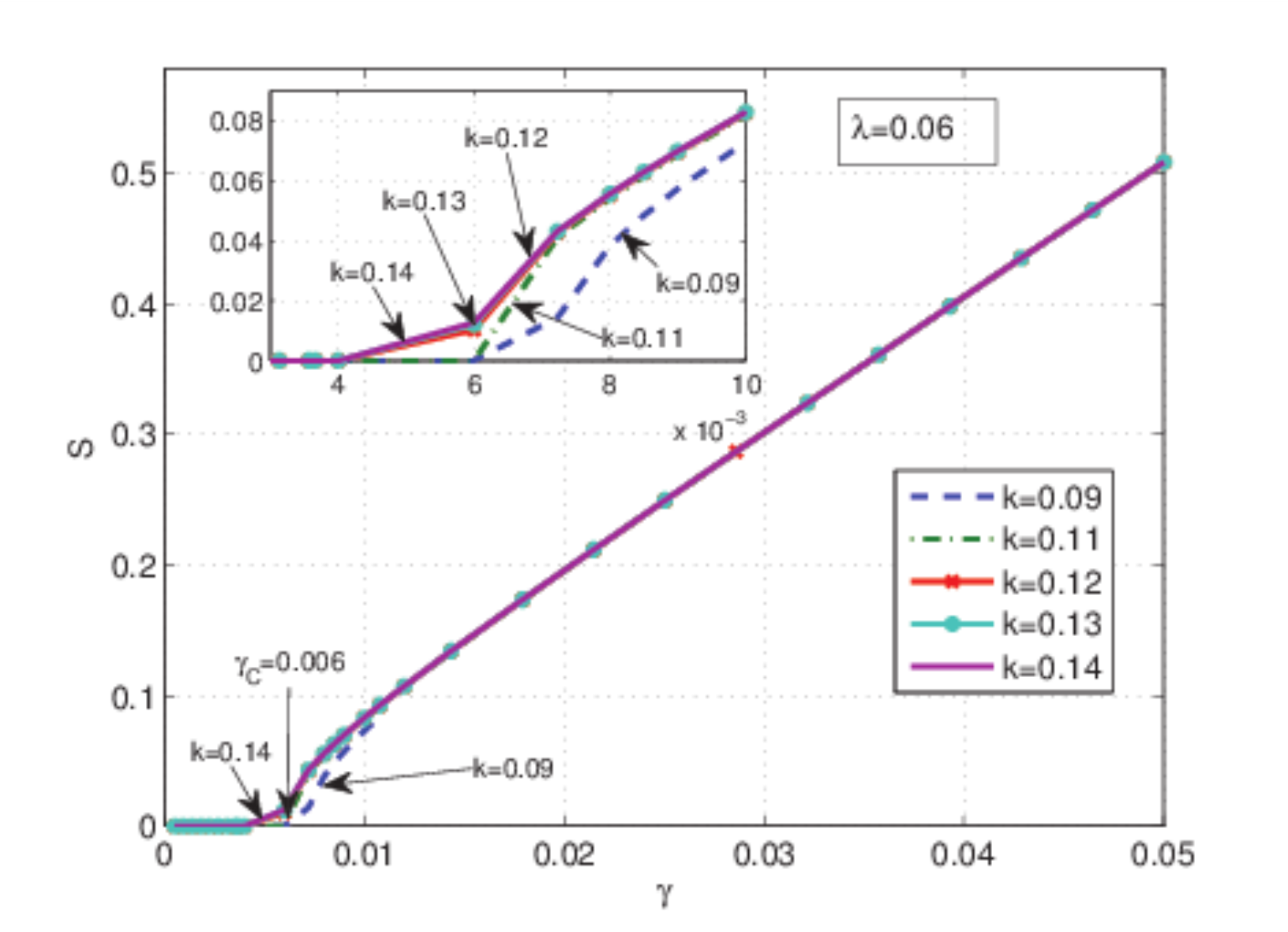}
\end{center}
\caption{(Color online) The same as in Fig. \protect\ref{Pic_Fig3}, but for $%
\protect\lambda =0.06$, and propagation constants $k=0.09$, $k=0.11$, $0.12$%
, $0.13$, $0.14$. The inset zooms in on a vicinity of the destabilization
transition.}
\label{Pic_Fig4}
\end{figure}

We studied the stability of the 2D $\mathcal{PT}$-symmetric solitons through
the calculation of eigenvalues, $\sigma $, for modes of small perturbations
governed by the linearization of Eqs. (\ref{Psi}) and (\ref{Phi}) around the
stationary solitons, using methods elaborated in Ref. \cite{Yang:2010a}. The
main results are summarized in Figs. \ref{Pic_Fig3} and \ref{Pic_Fig4},
which show the dependence of the largest instability growth rate, $S=\max
\left\{ \mathrm{Re}(\sigma )\right\} >0$, on $\gamma $ (in fact, the
unstable eigenvalues are complex ones), for two different values of the
coupling constant, $\lambda =0.08$ and $0.06$, and a set of fixed values of
the propagation constant from the ranges of $0.12\leq k\leq 0.15$ and $%
0.09\leq k\leq 0.14$, respectively.

The ground-state $\mathcal{PT}$-symmetric solitons are stable in the region
where $S(\gamma )=0$, i.e., at $\gamma <\gamma _{\mathrm{C}}\approx 0.008$
and $\gamma <\gamma _{\mathrm{C}}\approx 0.006$ in the former and latter
cases, respectively. These critical values are essentially smaller (roughly,
by a factor of $10$) than the respective maximum possible values of the
gain-loss coefficient, $\gamma _{\max }=\lambda $, see Eq. (\ref{cr}),
because the 2D solitons were taken at points which are close to the
symmetry-breaking instability threshold in the usual coupler model, with $%
\gamma =0$, cf. Ref. \cite{Dror:2011a}. A typical example of a stable
soliton is displayed in inset (a) to Fig. Fig. \ref{Pic_Fig3}.

The critical value, $\gamma _{\mathrm{C}}$, depends on $k$, as shown in
detail in inset (b) to Fig. \ref{Pic_Fig3}, and in the inset to Fig. \ref%
{Pic_Fig4}. Naturally, $\gamma _{\mathrm{C}}$ increases with the decrease of
$k$, as smaller $k$ correspond to a smaller energy of the soliton, pushing
it farther from the symmetry-breaking threshold. However, the computation
for $k$ essentially smaller than those presented in Figs. \ref{Pic_Fig3} and %
\ref{Pic_Fig4} is difficult, as the soliton becomes too broad, and cannot
fit to the domain employed for the numerical solution.

The predictions of the stability and instability, produced by the
computation of the eigenvalues for perturbation modes, were verified by
direct simulations of Eqs. (\ref{GPE-U})-(\ref{GPE-V}), which were carried
out by means of the split-step method \cite{CppRecipes2002}. The numerical
algorithm was set in the $\left( x,y\right) $ domain of size $30\times 30$
[in the same notation in which Eqs. (\ref{GPE-U}) and (\ref{GPE-V}) are
written] with periodic boundary conditions. The domain was covered by a
discretization mesh of $256\times 256$ points. Because the instability of
the soliton, if any, is caused by the breakup of the symmetry between the $%
\Psi $ and $\Phi $ components, which are subject to the action of the gain
and loss, respectively [see Eqs. (\ref{GPE-U}) and (\ref{GPE-V})], the
initial perturbation in the simulations was introduced by multiplying the
two components, severally, by the following factors:
\begin{equation}
\Psi _{0}\left( x,y\right) \rightarrow 1.03\times \Psi _{0}\left( x,y\right)
,~\Phi _{0}\left( x,y\right) \rightarrow ~0.97\times \Phi _{0}\left(
x,y\right) .  \label{pert}
\end{equation}

\begin{figure}[tbp]
\begin{center}
\includegraphics[scale=0.90]{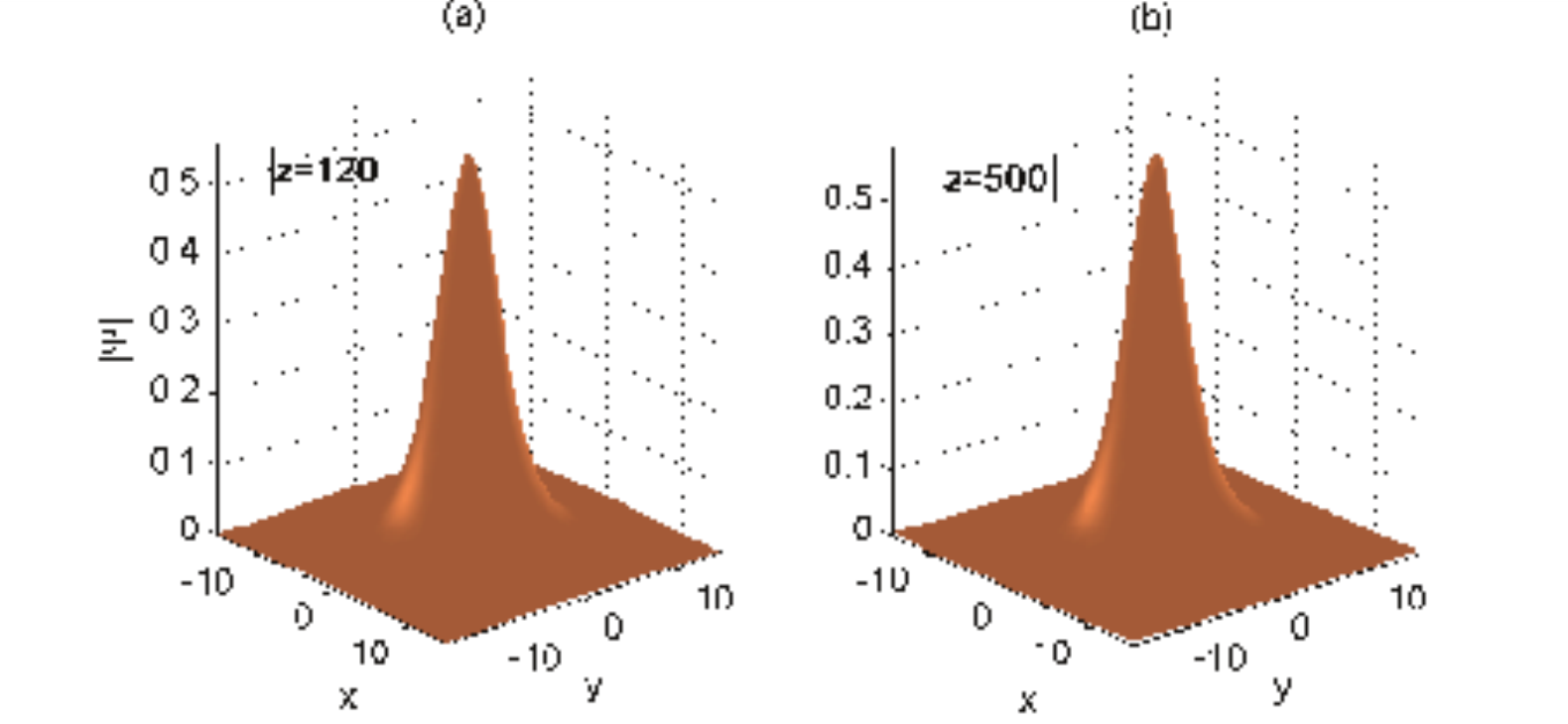}
\end{center}
\caption{(Color online) An example of the long-distance evolution of a
stable $\mathcal{PT}$-symmetric soliton, produced by direct simulations of
Eqs. (\protect\ref{GPE-U}) and (\protect\ref{GPE-V}) with perturbed initial
conditions [see Eq. (\protect\ref{pert})], starting from $z=0$. Panels (a)
and (b) show the shape of the $\Psi $ component at $z=120$ and $z=500$,
respectively. The evolution of the $\Phi $ component is similar. The
parameters are $k=0.12$, $\protect\lambda =0.08$, $\protect\gamma =0.002$.}
\label{Pic_Fig5}
\end{figure}

A typical example of the perturbed evolution of a stable soliton is
displayed in Fig. \ref{Pic_Fig5} [it is the same soliton whose stationary
shape is displayed in inset (a) of Fig. \ref{Pic_Fig3}]. The simulations
demonstrate the stability of the soliton in the course of the evolution over
the propagation distance which corresponds, roughly, to $20$ diffraction
lengths of the soliton (in fact, the simulations confirm the stability over
much longer distances).

\begin{figure}[tbp]
\begin{center}
\includegraphics[scale=0.75]{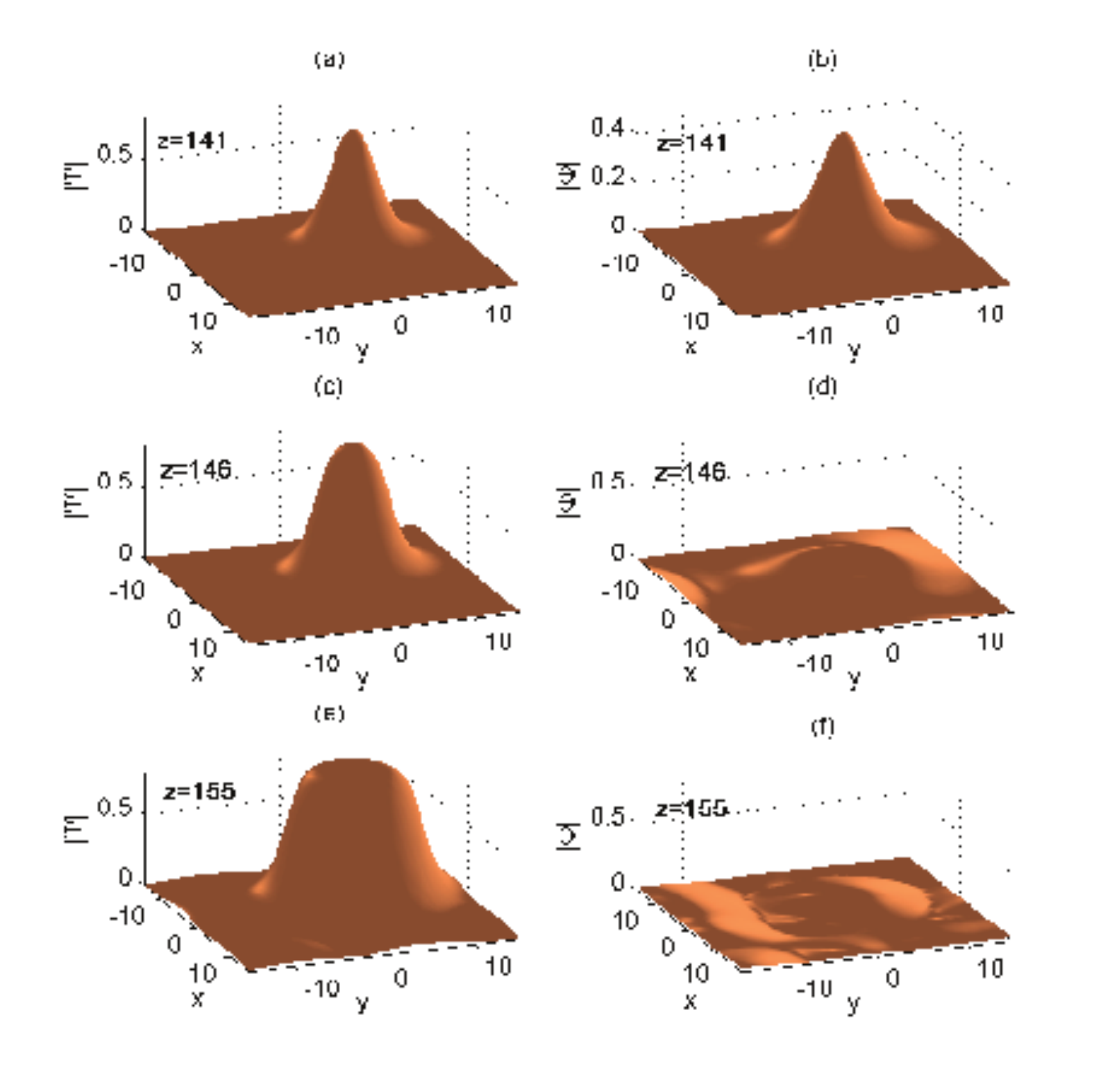}
\end{center}
\caption{(Color on line) The evolution of $\Psi $ and $\Phi $ components of
the unstable soliton (initiated at $z=0$) in the case of $k=0.12$, $\protect%
\lambda =0.08$ and $\protect\gamma =0.078>\protect\gamma _{\mathrm{C}}$. The
panels display the shapes of the two components at indicated values of the
propagation distance.}
\label{Pic_Fig6}
\end{figure}

An example of the unstable evolution, observed at $\gamma =0.078>\gamma _{%
\mathrm{C}}$, is displayed in Fig. \ref{Pic_Fig6}. The initial perturbation
was again introduced as per Eq. (\ref{pert}); without the perturbation, the
soliton may seem stable in the course of a long simulation, as the
instability is quite weak. As seen in in Fig. \ref{Pic_Fig6}, the
transmission over distance $z\simeq 150$, which is estimated as $\sim 6$
diffraction lengths, leads to destruction of the soliton. In fact, this
propagation distance is long, which once again stresses that we are here
dealing with weak instability. Moreover, it is relevant to note that the
value of $\gamma =0.078$, selected for this simulation, is very close to $%
\gamma _{\max }=\lambda =0.08,$ see Eq. (\ref{cr}). For smaller values of $%
\gamma $, Fig. \ref{Pic_Fig3} suggests that the propagation distance
necessary for the development of the instability, $z_{\mathrm{instab}}\sim
S^{-1}$, may be an order of magnitude larger than in Fig. \ref{Pic_Fig6}.
This conclusion suggests that, in terms of physical applications (such as
collisions between solitons, see \ below), unstable solitons may actually be
robust objects.

Getting back to the analysis of the instability development, Fig. \ref%
{Pic_Fig6} demonstrates, that, as may be expected, the $\Phi $
component of the unstable soliton decays into a vanishingly small
pattern under the action of the linear loss. On the other hand, the
pump of energy by the gain into the $\Psi $ component does not cause
an indefinite growth of its amplitude, but rather makes this
component progressively ``fatter" (broader). The latter result is
easily explained by the character of the 2D fundamental-soliton
solutions of the single NLS equation with the CQ nonlinearity: in
the limit of large energy, the amplitude of the soliton is
bounded by the largest value given by Eq. (\ref{4/3}), $U_{\max }=\sqrt{3}%
/2\approx \allowbreak 0.87$, which is consistent with Figs. \ref{Pic_Fig6}%
(c,e), while the effective radius of the ``fat" soliton, $R$%
, is related to its total energy as $E\approx \pi R^{2}U_{\max }^{2}=\left(
3/4\right) \pi R^{2}$. In the combination with the energy-balance equation (%
\ref{dE/dz}), this argument predicts that the radius of the unstable soliton
eventually grows exponentially, $R(z)\approx R_{0}\exp \left( \gamma
z\right) $.

\subsection{Collisions between solitons}

Because the underlying equations, (\ref{GPE-U}) and (\ref{GPE-V}), maintain
the Galilean invariance, in spite of the presence of the gain and loss terms
in them, the application of the kick to the soliton,
\begin{equation}
\left\{ \Psi ,\Phi \right\} \rightarrow \exp \left( i\mathbf{qr}\right)
\left\{ \Psi ,\Phi \right\} ,  \label{kick}
\end{equation}%
will set it in motion with velocity $\mathbf{V}=2\mathbf{q}$ [in fact, it is
a tilt in the $\left( x,y,z\right) $ space]. Then, comparison with the work
previously done for fundamental solitons in the single-core CQ model \cite%
{Sweden,interaction} suggests to consider collisions between moving (tilted)
2D solitons in the present \ model. A straightforward consideration of
physical parameters relevant for the optical waveguides demonstrates that $%
q\sim 1$ in the present notation corresponds to the tilt of the propagation
direction $\sim 0.1^{\mathrm{o}}$.

Strictly speaking, the collisions should be considered only between fully
stable solitons. However, it was shown above that those solitons which are
unstable may be subject to a very weak instability. The propagation distance
needed for simulating collisions between solitons is actually much smaller
than the above-mentioned instability distance. Therefore, the consideration
of collisions between\ weakly unstable solitons is meaningful too.

\begin{figure}[tbp]
\par
\begin{center}
\includegraphics[scale=0.75]{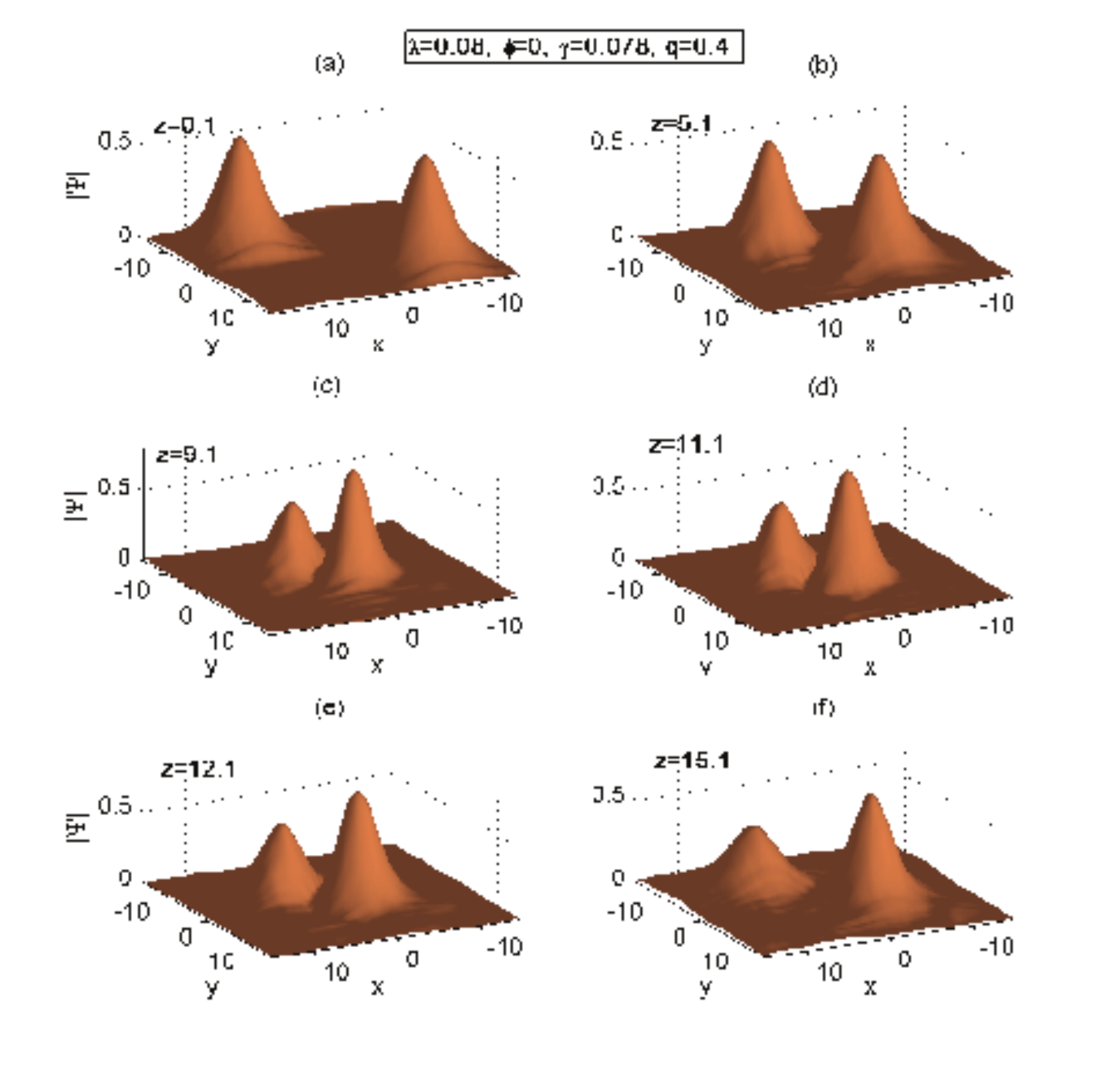}
\end{center}
\caption{(Color on line) Collision between two identical solitons with zero
phase shift, kicked as per Eq. (\protect\ref{kick}) in the opposite
directions by $q=0.4$. The panels display the shapes of the $\Psi $%
-components at indicated values of the propagation distance. The evolution
of the $\Phi $ component is the same, as the collision does not break the $%
\mathcal{PT}$ symmetry. The parameters are $k=0.12$, $\protect\lambda =0.08$%
, $\protect\gamma =0,078$.}
\label{Pic_Fig7}
\end{figure}

To simulate the collisions, two well-separated replicas of a stationary
soliton, with phase shift $\phi $ between them, were created and kicked in
opposite directions as per Eq. (\ref{kick}), i.e., with factors $\exp \left(
\pm i\mathbf{qr}\right) $, so as to initiate the head-on collision between
them. An example of the simulated collision between two identical in-phase ($%
\phi =0$) solitons with the same parameters as in Fig. \ref{Pic_Fig6} is
displayed in Fig. \ref{Pic_Fig7}, for kicks $q=0.4$. It is seen that the
collision is inelastic but not destructive. It gives rise to a spontaneous
symmetry breaking between the solitons, making one of them taller than the
other. The effect of the symmetry breaking between colliding solitons is
known in other models, see, e.g., Ref. \cite{Javid}. After the collision,
the two asymmetric solitons separate. The $\mathcal{PT}$ symmetry is kept
intact in the course of the collision (therefore, only the $\Phi $ component
is displayed in Fig. \ref{Pic_Fig6}), in spite of the fact that the
colliding solitons are classified, strictly speaking, as unstable ones. The
robustness of the setting against breaking of the $\mathcal{PT}$ symmetry is
explained by the fact that the collision happens over propagation distance $%
z\simeq 15$, which is ten times smaller than the distance necessary for the
manifestation of the instability, cf. Fig. \ref{Pic_Fig6}.

\begin{figure}[tbp]
\par
\begin{center}
\includegraphics[scale=0.75]{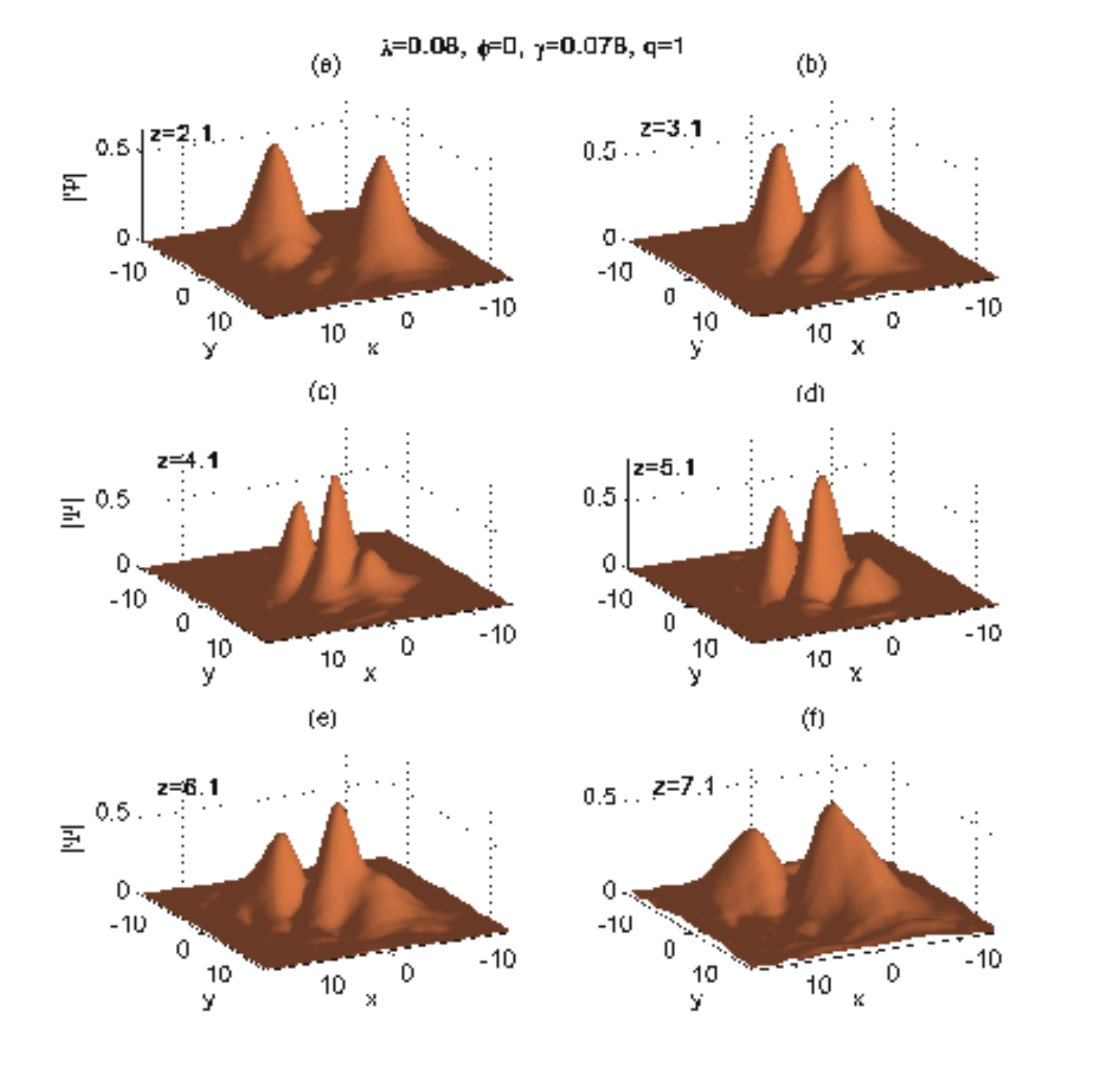}
\end{center}
\caption{(Color on line) The same as in Fig. \protect\ref{Pic_Fig7}, except
that the initial kick is larger, $q=1$.}
\label{Pic_Fig8}
\end{figure}

The collision induced by a larger kick, $q=1$, gives rise to a still
stronger effect of the spontaneously symmetry breaking between the two
solitons, as shown in Fig. \ref{Pic_Fig8}: one of the solitons temporarily
splits into two peaks of different heights, and later recombines back into a
single one. Nevertheless, in this case too, the collision does not break the
$\mathcal{PT}$ symmetry, which demonstrates the dynamical robustness of this
property.

We have also performed systematic simulations of collisions between fully
stable solitons, with $\gamma <\gamma _{\mathrm{C}}$, and also between
formally unstable or fully stable ones with the phase shift of $\phi =\pi $.
The results (not shown here in detail) demonstrate that truly stable
solitons collide in exactly the same fashion as their formally unstable
counterparts, i.e., like in Fig. \ref{Pic_Fig7} or Fig. \ref{Pic_Fig8} at
smaller and larger values of the kick, respectively. The introduction of the
phase shift $\phi =\pi $ does not change the results conspicuously either
(one may expect that the phase shift will be important in the case of very
small collision velocities \cite{Javid}).

\section{Conclusion}

We have considered 1D solitons (in a brief form), and their 2D counterparts
(systematically) in the model of the $\mathcal{PT}$-symmetric coupler, which
is characterized by the mutually balanced linear gain and loss applied to
its cores, in the combination with the CQ (cubic-quintic) intrinsic
nonlinearity acting in both cores. The self-defocusing quintic term is
necessary to protect the solitons against the usual 2D collapse. The model
can be realized in terms of the spatiotemporal transmission in dual-core
optical waveguides. The $\mathcal{PT}$-symmetric solitons lose their
stability with the increase of the energy, and restore the stability at
still larger energies. The boundary value of the linear-coupling strength,
above which the symmetric solitons are completely stable, was found by means
of an analytical approximation (using the exact solution for the CW version
of the system), which is rather close to its numerically found counterparts
for the 1D and 2D solitons. Stability limits of the 2D solitons and
evolution of the unstable ones were investigated in the numerical form, by
means of the computation of stability eigenvalues and direct simulations.
Although large parts of the soliton families are unstable, the instability
is quite weak, making it possible to use formally unstable solitons in
physical applications. Head-on collisions between solitons were studied in a
systematic form too, demonstrating that the collisions break the symmetry
between identical solitons, but do not break the $\mathcal{PT}$ symmetry.

It may be interesting to extend the 2D analysis of the same model to vortex
solitons. Starting from pioneering works \cite{Manolo} and \cite{Bob}, the
stability of 2D vortex solitons in the single-core system with the CQ
nonlinearity was a subject of many studies \cite{vort-CQ}. For the dual-core
CQ system, the analysis of the vortex-soliton stability was reported in Ref.
\cite{Dror:2011a}. The $\mathcal{PT}$-symmetric generalization of such
settings may be the next natural step.

\section*{Acknowledgments}

This work was partially supported by CONACyT (Mexico) grant No. 169496.
B.A.M. appreciates hospitality of Centro de Investigaci\'{o}n en Ingenier%
\'{\i}a y Ciencias Aplicadas at Universidad Aut\'{o}noma del Estado de
Morelos (Cuernavaca, Mexico).


\begin{thebibliography}{99}
\bibitem{Bender_review} C. M. Bender, Rep. Prog. Phys. \textbf{70}, 947
(2007).

\bibitem{special-issues} See special issues: H. Geyer, D. Heiss, and M.
Znojil, Eds., J. Phys. A: Math. Gen. \textbf{39}, \textit{Special Issue
Dedicated to the Physics of Non-Hermitian Operators} (\textit{PHHQP IV})
(University of Stellenbosch, South Africa, 2005) (2006); A. Fring, H. Jones,
and M. Znojil, Eds., J. Math. Phys. A: Math Theor. \textbf{41}, \textit{%
Papers Dedicated to the Subject of the 6th International Workshop on
Pseudo-Hermitian Hamiltonians in Quantum Physics} (\textit{PHHQPVI}) (City
University London, UK, 2007) (2008); C. Bender, A. Fring, U. G\"{u}nther,
and H. Jones, Eds., \textit{Special Issue: Quantum Physics with
non-Hermitian Operators}, J. Math. Phys. A: Math Theor. \textbf{41}, No. 44
(2012).

\bibitem{Muga} A. Ruschhaupt, F. Delgado, and J. G. Muga, J. Phys. A: Math.
Gen. \textbf{38}, L171 (2005); R. El-Ganainy, K. G. Makris, D. N.
Christodoulides, and Z. H. Musslimani, Opt. Lett. \textbf{32}, 2632 (2007);
M. V. Berry, J. Phys. A: Math. Theor. \textbf{41}, 244007 (2008).

\bibitem{experiment} A. Guo, G. J. Salamo, D. Duchesne, R. Morandotti, M.
Volatier-Ravat, V. Aimez, G. A. Siviloglou, and D. N. Christodoulides, Phys.
Rev. Lett. \textbf{103}, 093902 (2009); C. E. R\"{u}ter, K. G. Makris, R.
El-Ganainy, D. N. Christodoulides, M. Segev, and D. Kip, Nature Phys.
\textbf{6}, 192 (2010); A. Regensburger, C. Bersch, M.-A. Miri, G.
Onishchukov, D. N. Christodoulides, and U. Peschel, Nature \textbf{488}, 167
(2012).

\bibitem{PT_periodic} K. G. Makris, R. El-Ganainy, D. N. Christodoulides,
and Z. H. Musslimani, Phys. Rev. Lett. \textbf{100}, 103904 (2008); S.
Longhi, Phys. Rev. A \textbf{81}, 022102 (2010).

\bibitem{review} K. G. Makris, R. El-Ganainy, D. N. Christodoulides, and Z.
H. Musslimani, Int. J. Theor. Phys. \textbf{50}, 1019 (2011).


\bibitem{Musslimani2008} Z. H. Musslimani, K. G. Makris, R. El-Ganainy, and
D. N. Christodoulides, Phys. Rev. Lett. \textbf{100}, 030402 (2008); Z. Lin,
H. Ramezani, T. Eichelkraut, T. Kottos, H. Cao, and D. N. Christodoulides,
Phys. Rev. Lett. \textbf{106}, 213901 (2011); X. Zhu, H. Wang, L.-X. Zheng,
H. Li, and Y.-J. He, Opt. Lett. \textbf{36}, 2680 (2011); C. Li, H. Liu, and
L. Dong, Opt. Exp. \textbf{20}, 16823 (2012); C. M. Huang, C. Y. Li, and L.
W. Dong, \textit{ibid}. \textbf{21}, 3917 (2013).

\bibitem{Yang} S. Nixon, L. Ge, and J. Yang, Phys. Rev. A \textbf{85},
023822 (2012).

\bibitem{dark} H. G. Li, Z. W. Shi, X. J. Jiang, and X. Zhu, Opt. Lett.
\textbf{36}, 3290 (2011); V. Achilleos, P. G. Kevrekidis, D. J.
Frantzeskakis, and R. Carretero-Gonz\'{a}lez, Phys. Rev. A \textbf{86},
013808 (2012).

\bibitem{chi2} F. C. Moreira, F. K. Abdullaev, V. V. Konotop, and A. V.
Yulin, Phys. Rev. A \textbf{86}, 053815 (2012); F. C. Moreira, V. V.
Konotop, and B. A. Malomed, \textit{ibid}. \textbf{87}, 013832 (2013).

\bibitem{discrete} S. V. Dmitriev, A. A. Sukhorukov, and Y. S. Kivshar, Opt.
Lett. \textbf{35}, 2976 (2010); S. V. Suchkov, B. A. Malomed, S. V.
Dmitriev, and Y. S. Kivshar, Phys. Rev. E \textbf{84}, 046609 (2011); S. V.
Suchkov, A. A. Sukhorukov, S. V. Dmitriev, and Y. S. Kivshar, EPL \textbf{100%
}, 54003 (2012); D. A. Zezyulin and V. V. Konotop, Phys. Rev. Lett. \textbf{%
108}, 213906 (2012).

\bibitem{circular} D. Leykam, V. V. Konotop, and A. S. Desyatnikov, Opt.
Lett. \textbf{38}, 371 (2013); I. V. Barashenkov, L. Baker, and N. V.
Alexeeva, Phys. Rev. A \textbf{87}, 033819 (2013).

\bibitem{KPZ} K. Li and P. G. Kevrekidis, Phys. Rev. E \textbf{83}, 066608
(2011); V. V. Konotop, D. E. Pelinovsky, and D. A. Zezyulin, EPL \textbf{100}%
, 56006 (2012); J. D'Ambroise, P. G. Kevrekidis, and S. Lepri, J. Phys. A:
Math. Theor. \textbf{45}, 444012 (2012); K. Li, P. G. Kevrekidis, B. A.
Malomed, and U. G\"{u}nther, \textit{ibid}. \textbf{45}, 444021 (2012); M.
Kreibich, J. Main, H. Cartarius, and G. Wunner, Phys. Rev. A 87, 051601(R)
(2013).

\bibitem{PhysicaD} B. A. Malomed, Physica D \textbf{29}, 155 (1987); S.
Fauve, O. Thual, Phys. Rev. Lett. \textbf{64}, 282 (1990); W. van Saarloos
and P. C. Hohenberg, Phys. Rev. Lett. \textbf{64}, 749 (1990); V. Hakim, P.
Jakobsen, and Y. Pomeau, Europhys. Lett. 11, \textbf{19} (1990); B. A.
Malomed and A. A. Nepomnyashchy, Phys. Rev. A \textbf{42}, 6009 (1990); P.
Marcq, H. Chat\'{e}, and R. Conte, Physica D \textbf{73}, 305 (1994); T.
Kapitula and B. Sandstede, J. Opt. Soc. Am. B \textbf{15}, 2757 (1998); A.
Komarov, H. Leblond, and F. Sanchez, Phys. Rev. E \textbf{72}, 025604 (2005).

\bibitem{Kutz} J. N. Kutz, SIAM Rev. \textbf{48}, 629 (2006).

\bibitem{AKKZ} F. Kh. Abdullaev, Y. V. Kartashov, V. V. Konotop, and D. A.
Zezyulin, Phys. Rev. A \textbf{83}, 041805(R) {(2011)}; D. A. Zezyulin, Y.
V. Kartashov, V. V. Konotop, Europhys. Lett. \textbf{96}, 64003 (2011); A.
E. Miroshnichenko, B. A. Malomed, and Y. S. Kivshar, Phys. Rev. A \textbf{84}%
, 012123 (2011).

\bibitem{combined} Y. He, X. Zhu, D. Mihalache, J. Liu, and Z. Chen, Phys.
Rev. A \textbf{85}, 013831 (2012);
Opt. Commun. \textbf{285}, 3320 (2012).

\bibitem{Driben} R. Driben and B. A. Malomed, Opt. Lett. \textbf{36}, 4323
(2011).

\bibitem{Canberra} N. V. Alexeeva, I. V. Barashenkov, A. A. Sukhorukov, and
Y. S. Kivshar, Phys. Rev. A \textbf{85}, 063837 (2012).

\bibitem{Driben2} R. Driben and B. A. Malomed, Europhys. Lett. \textbf{94},
37011 (2011); I. V. Barashenkov, S. V. Suchkov, A. A. Sukhorukov, S. V.
Dmitriev, and Y. S. Kivshar, Phys. Rev. A \textbf{86}, 053809 (2012).

\bibitem{Konotop} F. K. Abdullaev, V. V. Konotop, M. \"{O}gren, and M. P. S%
\o rensen, Opt. Lett. \textbf{36}, 4566 (2011); Yu. V. Bludov, V. V.
Konotop, and B. A. Malomed, Phys. Rev. A \textbf{87}, 013816 (2013); Yu. V.
Bludov, R. Driben, V. V. Konotop, and B. A. Malomed, J. Opt. \textbf{15},
064010 (2013).

\bibitem{Dror:2011a} N. Dror and B. A. Malomed, Physica D \textbf{240}, 526
(2011).

\bibitem{Wabnitz} E. M. Wright, G. I. Stegeman, and S. Wabnitz, Phys. Rev. A
\textbf{40}, 4455 (1989); M. Romagnoli, S. Trillo, and S. Wabnitz, Opt. and
Quant. Elect. \textbf{24}, S1237 (1992).

\bibitem{coupler-Kerr} C. Par\'{e} and M. F\l orja\'{n}czyk, Phys. Rev. A
\textbf{41}, 6287 (1990); A. I. Maimistov, Kvantovaya Elektron. (Moscow)
\textbf{18}, 758 (1991) [English translation:, Sov. J. Quantum Electr.,
\textbf{21}, 687 (1991)]; N. Akhmediev and A. Ankiewicz, Phys. Rev. Lett.
\textbf{70}, 2395 (1993); J. M. Soto-Crespo and N. Akhmediev, Phys. Rev. E
\textbf{48}, 4710 (1993); P. L. Chu, B. A. Malomed, and G. D. Peng, J. Opt.
Soc. Am. B \textbf{10}, 1379 (1993); B. A. Malomed, I. Skinner, P. L. Chu,
and G. D. Peng, Phys. Rev. E \textbf{53}, 4084 (1996); W. C. K. Mak, B. A.
Malomed, and P. L. Chu, J. Opt. Soc. Am. B \textbf{15}, 1685 (1998).

\bibitem{coupler-chi2} W. C. K. Mak, B. A. Malomed and P. L. Chu, Phys. Rev.
E \textbf{55}, 6134 (1997); \textit{ibid}. \textbf{57}, 1092 (1998).

\bibitem{book} \textit{Spontaneous Symmetry Breaking, Self-Trapping, and
Josephson Oscillations}, B. A. Malomed, editor (\noindent Springer-Verlag:
Berlin and Heidelberg, 2013).

\bibitem{JYang} J. Yang, Stud. Appl. Math., in press.

\bibitem{collapse} L. Berg\'{e}, Phys. Rep. \textbf{303}, 259 (1998); E. A.
Kuznetsov and F. Dias, \textit{ibid}. \textbf{507}, 43 (2011).

\bibitem{CQ} G. S. Agarwal and S. Dutta Gupta, Phys. Rev. A \textbf{38},
5678 (1988); F. Smektala, C. Quemard, V. Couderc, A. Barth\'{e}l\'{e}my, J.
Non-Cryst. Solids \textbf{274}, 232 (2000); C. Zhan, D. Zhang, D. Zhu, D.
Wang, Y. Li, D. Li, Z. Lu, L. Zhao, Y. Nie, J. Opt. Soc. Am. B \textbf{19},
369 (2002); G. Boudebs, S. Cherukulappurath, H. Leblond, J. Troles, F.
Smektala, F. Sanchez, Opt. Commun. \textbf{219}, 427 (2003); F. Sanchez, G.
Boudebs, S. Cherukulappurath, H. Leblond, J. Troles, F. Smektala, J. Nonlin.
Opt. Phys. \& Mat. \textbf{13}, 7 (2004); K. Ogusu, J. Yamasaki, S. Maeda,
M. Kitao, M. Minakata, Opt. Lett. \textbf{29}, 265 (2004); E. L. Falc\~{a}%
o-Filho, C. B. de Ara\'{u}jo, and J. J. Rodrigues, Jr., J. Opt. Soc. Am. B
\textbf{24}, 2948 (2007).

\bibitem{review-Wise} B. A. Malomed, D. Mihalache, F. Wise, and L. Torner,
J. Opt. B: Quant. Semicl. Opt. \textbf{7}, R53 (2005).

\bibitem{Albuch} L. Albuch and B. A. Malomed, Mathematics and Computers in
Simulation \textbf{74}, 312 (2007).

\bibitem{Zeev} Z. Birnbaum and B. A. Malomed, Physica D \textbf{237}, 3252
(2008).

\bibitem{Lazar} L. Gubeskys and B. A. Malomed, J. Opt. Soc. Am. B \textbf{30}%
, 1843 (2013).

\bibitem{Bulgaria} Kh. I. Pushkarov, D. I. Pushkarov, and I. V. Tomov, Opt.
Quant. Electr. \textbf{11}, 471 (1979); S. Cowan, R. H. Enns, S. S.
Rangnekar, S. S. Sanghera, Can. J. Phys. \textbf{64}, 311 (1986).

\bibitem{Sweden} M. L. Quiroga-Teixeiro, A. Berntson, and H. Michinel, J.
Opt. Soc. Am. B \textbf{16}, 1697 (1999).

\bibitem{surface-tension} D. Novoa, H. Michinel, and D. Tommasini, Phys.
Rev. Lett. \textbf{103}, 023903 (2009).

\bibitem{Yang:2010a} J. Yang, \textit{Nonlinear Waves in Integrable and
Non-integrable Systems} (SIAM: Philadelphia, 2010).


\bibitem{CppRecipes2002} W. H. Press, S. A. Teukovsky, W. T. Vetterling, and
B. P. Flannery, \textit{Numerical recipes in C++} , Chap. 17 (Cambridge
University Press: Cambridge, 2002).

\bibitem{interaction} S. Konar, S. Jana, and M. Mishra, Opt. Commun. \textbf{%
255}, 114 (2005).

\bibitem{Javid} J. Atai and B. A. Malomed, Phys. Rev. E \textbf{64}, 066617
(2001).

\bibitem{Manolo} M. Quiroga-Teixeiro and H. Michinel, J. Opt. Soc. Am. B
\textbf{14}, 2004 1997).

\bibitem{Bob} R. L. Pego and H. A. Warchall, J. Nonlin. Sci. \textbf{12},
347 (2002).

\bibitem{vort-CQ} I. Towers, A. V. Buryak, R. A. Sammut, B. A. Malomed, L.
C. Crasovan, and D. Mihalache, Phys. Lett. A \textbf{288}, 292 (2001); B. A.
Malomed, L.-C. Crasovan, and D. Mihalache, Physica D \textbf{161}, 187
(2002); T. A. Davydova and A. I. Yakimenko, J. Opt. A: Pure Appl. Opt.
\textbf{6}, S197 (2004); H. Michinel, J. R. Salgueiro, and M. J. Paz-Alonso,
Phys. Rev. E \textbf{70}, 066605 (2004); M. J. Paz-Alonso and H. Michinel,
Phys. Rev. Lett. \textbf{94}, 093901 (2005); L. Dong, J. Wang, H. Wang, and
G. Yin, Phys. Rev. A \textbf{79}, 013807 (2009).
\end{thebibliography}
\end{document}